\documentclass[aps,prb,superscriptaddress,showkeys,showpacs,twocolumn,floatfix]{revtex4}

\usepackage{amssymb,amsbsy,amsmath}

\usepackage{graphicx}
\usepackage{dcolumn}
\usepackage{bm}
\usepackage{subfigure}

\usepackage{multirow}
\usepackage{array}
\usepackage{rotating}

\newcommand{\ket}[1]{\ensuremath{|#1\rangle}}

\newcommand{\op}[1]{%
    \fontdimen12\textfont3=2pt\fontdimen12\scriptfont3=1.4pt%
    \!\null\mathop{\vphantom{#1}\smash{#1}}\limits_{\sim}\null\!}

\newcommand{\fmref}[1]{(\protect\ref{#1})}
\newcommand{\xref}[1]{\protect\ref{#1}}
\newcommand{\figref}[1]{Fig.~\protect\ref{#1}}

\newcommand {\mofe} {\{$\textrm{Mo}_{72}\textrm{Fe}_{30}$\}}
\newcommand {\mocr} {\{$\textrm{Mo}_{72}\textrm{Cr}_{30}$\}}
\newcommand {\mov} {\{$\textrm{Mo}_{72}\textrm{V}_{30}$\}}
\newcommand {\wv} {\{$\textrm{W}_{72}\textrm{V}_{30}$\}}
\newcommand {\movsaw} {\{$\textrm{Mo}_{75}\textrm{V}_{20}$\}}

\begin{document}

\title{Determination of exchange energies in the sawtooth spin ring {\movsaw} by ESR}

\author{Yugo Oshima}
\email{yugo@riken.jp}
\affiliation{Institute for Materials Research, Tohoku University, Katahira 2-1-1, Sendai 980-8577, Japan}
\affiliation{RIKEN, Wako, Saitama 351-0198, Japan}
\author{Hiroyuki Nojiri}
\email{nojiri@imr.tohoku.ac.jp}
\affiliation{Institute for Materials Research, Tohoku University, Katahira 2-1-1, Sendai 980-8577, Japan}
\author{J\"urgen Schnack}
\affiliation{Fakult\"at f\"ur Physik, Universit\"at Bielefeld, Postfach 100131, D-33501 Bielefeld, Germany}
\author{Paul K\"ogerler}
\affiliation{Institut f{\"u}r Anorganische Chemie, RWTH Aachen, Landoltweg 1, D-52074 Aachen, Germany}
\author{Marshall Luban}
\affiliation{Ames Laboratory \& Department of Physics and Astronomy, Iowa State University, Ames, Iowa 50011, USA}

\date{\today}

\begin{abstract}
The magnetism of the polyoxometalate cluster {\movsaw}, containing a
sawtooth ring of 10 corner-sharing triangles located on the
equator of the barrel-shaped molecule, has remained debatable since it is masked
by contributions from impurities as well as
temperature-independent paramagnetism. In this article we
 demonstrate the usefulness of ESR measurements since the
temperature dependence of the ESR intensity can discriminate
between impurity and molecular contributions. We determine the
exchange parameters and therefore also the low-lying spectrum of {\movsaw}, 
especially the low-lying singlet states which so far have been
probed solely by specific heat measurements.
\end{abstract}


\maketitle

\section{Introduction}
\label{sec-1}

The series of Keplerate molecules {\mofe}, {\mocr}, {\mov} and {\wv} 
is one of the 
beautiful creations made possible by recent developments of
modern
chemistry.\cite{MPP:CR98,MSS:ACIE99,MKD:CCR01,MLS:CPC01,Mue:Sci03,MTS:AC05,BKH:CC05,KMS:CCR09,TMB:CC09,KTM:DT10}
In these nanosized molybdate or tungstate-based molecules, 
30 magnetic ions are located on the 30 vertices of an
icosidodecahedron and are antiferromagnetically coupled, resulting in  
triangular and pentagonal networks. As a result, this leads to a 
strongly frustrated cluster with a huge
number of quantum states.\cite{Sch:DT10} Since the number of
states, $(2s+1)^{30}$, can be varied by substituting the
magnetic ions ($2^{30}$ and $6^{30}$ for V $(s=1/2)$ and Fe
$(s=5/2)$, respectively), these are ideal systems for studying the
transitions from quantum to classical behavior.

\begin{figure}[ht!]
\centering
\includegraphics*[clip,width=75mm]{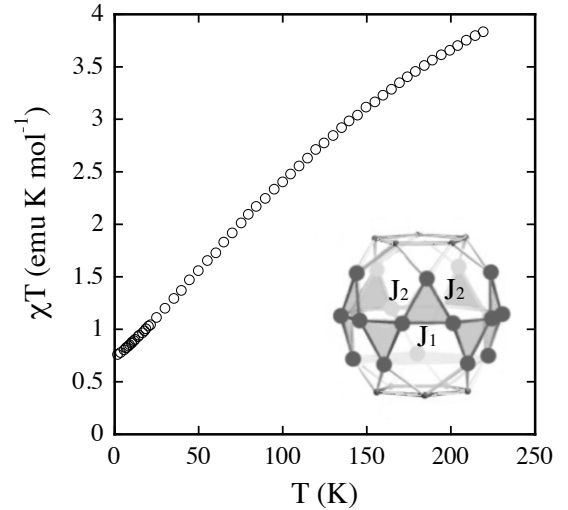}
\caption{Temperature dependence of $\chi T$ from
  Ref.~\onlinecite{MKD:CCR01}. 
  The inset shows a scheme of the metal skeleton of {\movsaw} highlighting the nearest-neighbor 
  exchange pattern between the V$^{4+}$ ions.}   
\label{v20-fig-1}
\end{figure}

Among synthesized derivatives of the Keplerate clusters, 
the compound {\movsaw} is analogous to \mov\ except that the 10
V$^{4+}$ ions $(s=1/2)$ located adjacent to the north and south poles of the icosidodecahedron are
substituted by nonmagnetic ions.\cite{MKB:CC99} 
Therefore, this compound is
equivalent to a sawtooth chain of 10 triangles with a
periodic boundary condition as schematically shown in the
inset of \figref{v20-fig-1}. 
Basically, {\movsaw} is classified as a sawtooth spin ring rather than a Keplerate cluster 
since it has only a partial substructure of {\mov}. 
However, determining the exchange couplings of {\movsaw} can be a good reference for the 
charactarization of its analogous compound {\mov}.  
Moreover, sawtooth chains also belong to the
class of frustrated antiferromagnetic spin systems with
potentially very unusual magnetization
curves,\cite{SHS:PRL02,RSH:JPCM04} as partly realized in the
recently investigated magnetic material
azurite.\cite{RWS:PRL08,HHP:JPCM11} 

For {\movsaw}, it turns out 
that the product, $\chi T$, of the magnetic susceptibility and temperature 
decreases steadily with decreasing temperature due to
antiferromagnetic couplings, and it was assumed in Ref.~\onlinecite{MKD:CCR01} that
the ground state is a singlet. Additionally, those authors 
estimated the exchange couplings between the magnetic ions from
a theoretical fit to a reduced model that contains only a ring
of six triangles. 
They concluded that their experimental results
are well reproduced when $J_1=288$~K, $J_2=0.55\cdot J_1$,
compare inset of \figref{v20-fig-1}, together with an additional
coupling $J_3=0.20\cdot J_1$ between the tips of every second
triangle.\cite{MKD:CCR01}

Takemura and Fukumoto refined those parameters by using a
finite-temperature Lanczos method for the full system of 10
coupled triangles.\cite{TaF:JPSJ07} They found similar exchange
coupling parameters, which are $J_1=388$~K, $J_2=0.42\cdot J_1$,
and $J_3=0.21\cdot J_1$. However, these values are still under
discussion for various reasons. The main concern is the
unrealistically large exchange $J_3$ which should act across a
distance of 11.7~\AA, mediated by multi-center exchange pathways. 
Another problem is given by the fact that
the susceptibility data are superimposed by an unknown
amount of free vanadium ions (impurities) as well as by
temperature-independent (i.e. van Vleck) paramagnetism typical of polyoxometalates.

We propose in this article that ESR is a unique
method that overcomes these difficulties. It can separate the
signals stemming from the intrinsic \movsaw\ and from impurities
since each ESR linewidth is qualitatively different. This
enables us to obtain the pure magnetic response from \movsaw\ in
contrast to the susceptibility measurements in which the
intrinsic and extrinsic responses are mixed. Since the
temperature dependence of the ESR intensity is related to the
energies of the excited states, we are able to obtain
information about the exchange couplings. The ratio of $J_1$
and $J_2$ determines the frustration of the sawtooth chain of
which the density of low-lying singlet states is a
fingerprint.\cite{WEB:EPJB98,BAA:PRL03} Usually the density of
singlet-levels is deduced from specific heat measurements, but
we will show below that ESR is also capable of determining the
density of low-lying singlet states.

The article is organized as follows. In Section~\xref{sec-2} we
briefly explain the experimental method. Section~\xref{sec-3}
contains our experimental results that are compared to the
calculated ESR intensities assuming various sets of model
parameters. The article closes with a short summary.

\section{Experimental}
\label{sec-2}

The Terahertz Electron Spin Resonance Apparatus in the Institute
for Materials Research (TESRA-IMR) of Tohoku University has been
used for the high field ESR measurements.\cite{NAA:NJP06} A
simple transmission method with Faraday configuration has been
employed. We used conventional Gunn oscillators for the
millimeter wave radiation, and an InSb detector for transmission
detection. In addition, a pulsed magnetic field up to 8.5~T can
be generated from a 90~kJ capacitor bank. Two types of
cryostats, a conventional $^4$He bath type cryostat and gas-flow
type cryostat, were used for the measurements for the low and high
temperature ranges, respectively. Powder samples were used in
this study.

\begin{figure}[ht!]
\centering
\includegraphics*[clip,width=75mm]{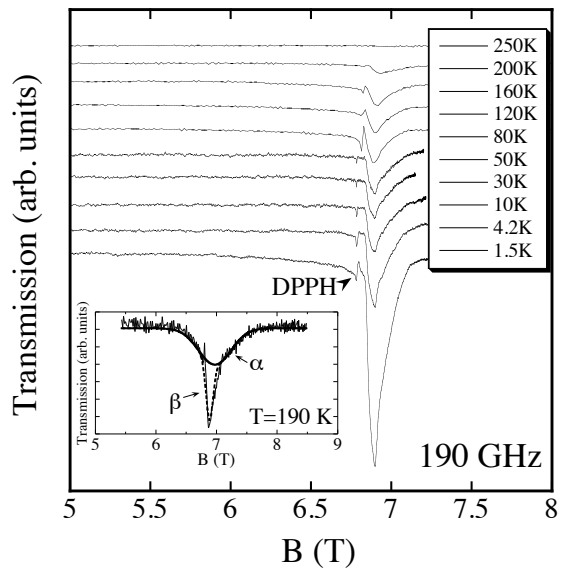}
\caption{Temperature dependence of ESR spectra for 190~GHz. The
very narrow peak observed around 6.8~T is a field marker. The inset
shows the spectrum for 190~K. The solid and broken lines are
fitting curves for resonance $\alpha$ and $\beta$, respectively
(see text for details).}
\label{v20-fig-2}
\end{figure}

\section{Results and discussion}
\label{sec-3}

\subsection{High field ESR results}

Figure~\xref{v20-fig-2} shows the typical temperature dependence
of the ESR spectra for \movsaw. The employed frequency is
190~GHz, and the temperature is varied from 1.5 to 250~K. As
shown in the inset of \figref{v20-fig-2}, two absorption lines,
a broad and a sharp one, are clearly observed at
190~K. Hereafter, we denote the observed resonances as $\alpha$
and $\beta$ for the former and the latter, respectively. The
tiny absorption observed at around 6.8~T is from DPPH, which stands for (2,2-diphenyl-1-picrylhydrazyl), 
a field marker. The intensity of $\alpha$ gradually increases by
decreasing the temperature, but quickly diminishes below 80~K,
which is a typical ESR behavior of the excited states. On the
other hand, the intensity of $\beta$ is inversely proportional
to the temperature, following Curie's law. Therefore, $\alpha$
can be assigned to the resonance originating from \movsaw, and
$\beta$ from the impurities at the cation sites.\cite{MKD:CCR01}
Since the singlet ground state is ESR silent, the absence of intrinsic ESR signal from {\movsaw} at low temperature and 
the ESR observation of the excited states for relatively high
temperatures suggests that the ground state of {\movsaw} is a
singlet, which is consistent with magnetic susceptibility results.\cite{MKD:CCR01}

\begin{figure}[ht!]
\centering
\includegraphics*[clip,width=75mm]{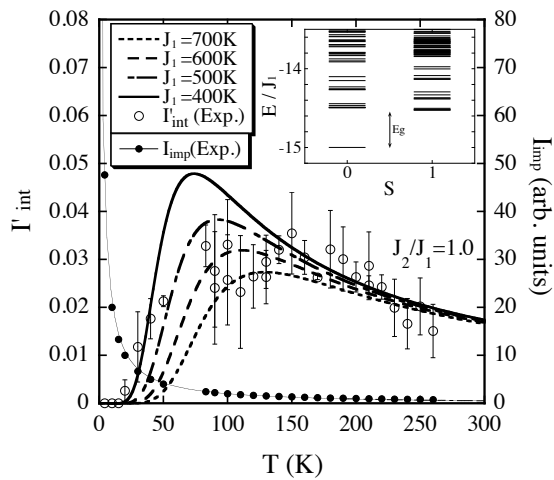}
\caption{Solid circles are the integrated intensities for
  resonance $\beta$ coming from paramagnetic impurities; they
  are fitted using equation \eqref{E-3-1} (thin
  solid curve). Open circles are the normalized integrated
  intensities for resonance $\alpha$ coming from the intrinsic
  \movsaw. Thick solid and dashed curves are calculated curves
  of the normalized integrated intensity for various values of
  $J_1$. The calculated curves are obtained from the of singlet
  and triplet energy levels (inset) assuming an equilateral
  triangle model. The energy is normalized by $J_1$.}
\label{v20-fig-3}
\end{figure}

In the inset of \figref{v20-fig-2} the absorption lines of
$\alpha$ and $\beta$ are fitted by Gaussian curves (solid and
broken curves, respectively), and integrated intensities for
each temperature are obtained. The integrated intensity of
impurities (i.e. of resonance $\beta$), $I_{\text{imp}}$, versus
temperature is shown as solid circles in
\figref{v20-fig-3}. According to the Boltzmann distribution, the
ESR intensity of the $s=1/2$ impurities can be written as
\begin{eqnarray}
\label{E-3-1}
I_{\text{imp}}
&=&
w\, N_{\text{imp}}\, \text{tanh}
\left(
\beta g \mu_B B/2
\right)
\ .
\end{eqnarray}
$w$ is a coefficient which is proportional to the power and
frequency of the radiation, $N_{\text{imp}}$ is the number of
impurities,\cite{AbB:DP86} and $\beta=1/k_B T$ the inverse
temperature.  $g$ is the spectroscopic splitting factor and
$\mu_B$ the Bohr magneton, respectively. By using equation
\fmref{E-3-1} $w N_{\text{imp}}=83.9\pm7.4$ is obtained from the
fitting curve in \figref{v20-fig-3} (thin solid curve).

On the other hand, the intrinsic ESR intensity of the excited
triplet states, $I_{\text{int}}$, which are the transitions
$\ket{M=-1}\rightarrow\ket{M=0}$ and
$\ket{M=0}\rightarrow\ket{M=+1}$, is proportional to the
radiation coefficient $w$ and the difference of the population
between the transition levels, i.e. 
\begin{eqnarray}
\label{E-3-2}
I_{\text{int}}
=
w\,
&\Big\{&
2
(N_{\text{int, M=-1}}-N_{\text{int, M=0}})
\\
&+&
2
(N_{\text{int, M=0}}-N_{\text{int, M=+1}})
\Big\}
\nonumber
\ .
\end{eqnarray}
The factor of 2 corresponds to the square of the transition
matrix element.\cite{AbB:DP86} When the intrinsic intensity
$I_{\text{int}}$ (i.e. intensity of $\alpha$) is normalized by $w
N_{\text{imp}}$, which was already obtained above, the unknown
radiation coefficient $w$ can be eliminated, and then we obtain
\begin{eqnarray}
\label{E-3-3}
I_{\text{int}}^\prime
&=&
\frac{N_{\text{int}}}{N_{\text{imp}}}
\times
\frac{2}{Z(T,B)}
\\
&&
\times
\sum_{\text{triplets}\ i}
\left\{
e^{-\beta (E_i-g\mu_B B)}
-
e^{-\beta (E_i+g\mu_B B)}
\right\}
\nonumber
\ ,
\end{eqnarray}
where $Z(T,B)$ is the partition function
\begin{eqnarray}
\label{E-3-4}
Z(T,B)
&=&
\sum_{\text{singlets}\ k}
e^{-\beta E_k}
+
\sum_{\text{triplets}\ i}
\\
&&
\left\{
e^{-\beta (E_i-g\mu_B B)}
+
e^{-\beta E_i}
+
e^{-\beta (E_i+g\mu_B B)}
\right\}
\nonumber
\ .
\end{eqnarray}
$N_{\text{int}}$ is the number of \movsaw, and $E_k$ and $E_i$
are the energies of the singlet and triplet states,
respectively. The normalized values of $I_{\text{int}}^\prime$
are presented as open circles in \figref{v20-fig-3}. The error
bars are obtained from the uncertainty of the fitting curves.
Using \eqref{E-3-3} $I_{\text{int}}^\prime$ can be evaluated
exactly using the energy levels of the singlet and triplet
states which can be obtained from the model Hamiltonian by
diagonalization. This way the exchange constants can be
determined by comparing experimental and theoretical values of
$I_{\text{int}}^\prime$.

It is important to note that the whole procedure works with only
singlet and triplet levels since the exchange constants are large
and therefore states with $S\ge 2$ are irrelevant.

\subsection{Theoretical Models and analysis}

The magnetism of \movsaw\ is modeled by a Heisenberg Hamiltonian
augmented with a Zeeman term
\begin{eqnarray}
\label{E-3-5}
H
&=&
\sum_{i<j}\;
J_{ij}
\op{\vec{s}}_i \cdot \op{\vec{s}}_{j}
+
g\, \mu_B\, B\,
\sum_{j}\;
\op{s}^z_j
\ .
\end{eqnarray}
Here $J_{ij}$ is the exchange interaction between spins at sites
$i$ and $j$. The eigenvalues of this Hamiltonian can either be
determined by complete matrix diagonalization or for the
low-lying levels by the Lanczos
method.\cite{Lan:JRNBS50,SHS:JCP07,ScS:IRPC10} For our
simulation we considered the coupling $J_1$ on the equator and
$J_2$ to the tips of the triangles as schematically shown in the
inset of \figref{v20-fig-1}. The low-energy part of the singlet
and triplet sectors for $J_2=J_1$ (i.e. equilateral triangle) is
shown in the inset of \figref{v20-fig-3}. Note that energies are
given as multiples of $J_1$. The lowest singlet level is
degenerate due the frustration of the triangular magnetic
structure.\cite{Sch:DT10}

From the energy eigenvalues we can calculate
$I_{\text{int}}^\prime$. The ratio
$N_{\text{int}}/N_{\text{imp}}=1/2$ is used since impurities of
about $2\mu_B$ per molecular unit are observed in the
magnetization measurements. It is also important to note that we
have taken into account all singlet and triplet levels up to $10
E_g$, where $E_g$ is the gap between the ground state and the
lowest triplet level (see inset of \figref{v20-fig-3}). In view
of the experimental temperature range this is more than
sufficient since the exchange parameters turn out to be rather
large.

The calculated $I_{\text{int}}^\prime$ curves for various
$J_1=J_2$ are shown in \figref{v20-fig-3}. The position of the
maximum of the calculated curves changes by varying the exchange
coupling constant. The same holds true for the initial rise of
the curve which is related to the energy gap between the singlet
and triplet states, i.e. $E_g$. On the other hand, the
high-temperature behavior of $I_{\text{int}}^\prime$ is in part
related to the number of low-lying singlet
states,\cite{SSL:PRB01} therefore, the high-temperature tail is
almost independent of the exchange parameters.  For
$J_1=J_2=400$~K (thick solid curve in \figref{v20-fig-3}), the
calculated curve fits well with the initial rising part and the
tail part. However, a large difference is seen on the peak
position. This suggests that an equilateral triangle model
($J_2=J_1$) is not suited for this system, and that an isosceles
triangle model ($J_2\neq J_1$) should be considered.

\begin{figure}[ht!]
\centering
\includegraphics*[clip,width=75mm]{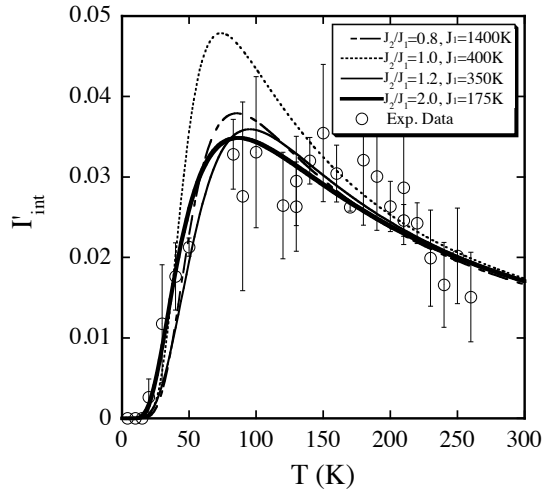}
\caption{Normalized integrated intensities for resonance
  $\alpha$ coming from the intrinsic \movsaw\ for various ratios
  $J_2/J_1$. The respective exchange parameters are shown in the
  legend.}
\label{v20-fig-4}
\end{figure}


Hence, we have investigated chains of isosceles triangles with various ratios $J_2/J_1$. 
Figure~\xref{v20-fig-4} shows the curves for 
$J_2/J_1=0.8, 1.2, 2.0$. The absolute values of the data are determined by
the low-temperature behavior, i.e. the singlet-triplet gap. It
is obvious that a much improved fit to our experimental data is provided by 
assuming the isosceles triangle model. The agreement in
the observable $I_{\text{int}}^\prime$ appears to be better if
$J_2>J_1$. We emphasize that there is no need
to introduce the additional exchange coupling $J_3$ between tips
of neighboring triangles. This exchange is very unlikely anyway
due to the very long exchange pathway.

\begin{figure}[ht!]
\centering
\includegraphics*[clip,width=75mm]{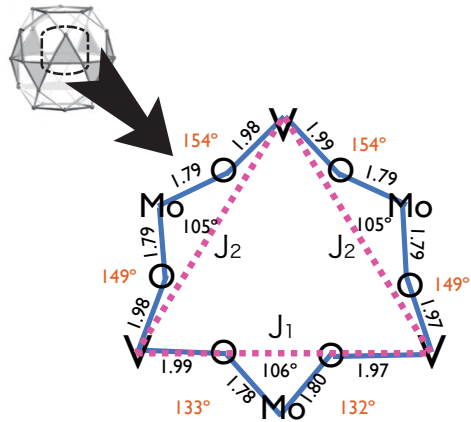}
\caption{Bond lengths and angles for an individual triangle of
  \movsaw.} 
\label{v20-fig-5}
\end{figure}

In order to rationalize the differences between $J_1$ and $J_2$
it is helpful to look at the crystal structure of an individual
triangle of \movsaw\ as schematically shown in
\figref{v20-fig-5}. While the bonding length between vanadium
atoms along $J_1$ and $J_2$ bonds does not vary much, the V-O-Mo
bonding angles do. Moreover, since the V-O-Mo bonding angles are
larger along $J_2$ bonds the Goodenough-Kanamori
rule\cite{Goo:PR55} suggests that the exchange should be
stronger along this pathway, which is in accord with our
observations.  Among the two good fits, $J_2/J_1=1.2, J_1=350$~K
and $J_2/J_1=2.0, J_1=175$~K, we tend to favor the second
parameter set since this reproduces the low-temperature behavior
much better. The absolute numbers are in good agreement with
other polyoxometalates containing V$^{4+}$ spin centers.\cite{MTS:AC05,BKH:CC05,TMB:CC09}

Our results suggest that the type of frustration which is
present in \movsaw\ is different from that in the original
Keplerate molecules which are akin to the kagome lattice. Not
only is \movsaw\ a quasi one-dimensional object, in addition the
magnetic centers are not equivalent. The special ratio of
$J_2/J_1=2.0$ relates it strongly to sawtooth chains with flat
bands of one-magnon energies.\cite{SHS:PRL02,RSH:JPCM04} Such
chains are characterized by giant magnetization steps of 50~\%
of the saturation magnetization. Unfortunately, the exchange
interactions present in \movsaw\ are much too large in order to observe such magnetization steps 
experimentally, even for the first one, let alone subsequent steps at larger fields.

\section{Summary}

In summary, we have performed ESR measurements of the polyoxometalate cluster {\movsaw}. 
We succeeded in separating the
intrinsic ESR signal of the molecules and the signal of the
magnetic impurities. Via the temperature dependence of the
integrated intensity we could deduce the parameters of the
underlying Heisenberg Hamiltonian. 
We also found that the additional exchange coupling $J_3$, 
which has been taken into account in previous studies \cite{MKD:CCR01,TaF:JPSJ07}, is not necessary to 
explain the magnetic properties of {\movsaw}.

Finally, we would like to
stress that ESR is a powerful tool to study the low energy
spectrum especially of frustrated magnetic systems. Since the
sensitivity of ESR is very high, this method can in certain
cases be far more effective than the specific heat
measurements in characterizing the low-lying density of
states. This study shows only ESR results for a single frequency
of 190~GHz. It is of course possible to obtain more detailed
information on the spectrum by combining results using multiple
frequencies.

\section*{Acknowledgment}

  This work was supported by the Deutsche Forschungsgemeinschaft
  through the research unit 945.  Ames Laboratory is operated
  for the U.S. Department of Energy by Iowa State University
  under Contract No.  W-7405-Eng-82. H.~N.  acknowledges the
  support by Grant in Aid for Scientific Research on Priority
  Areas (No.  13130204) from MEXT, Japan and by Shimazu Science
  Foundation.

\bibliography{/home/schnack/tex/bibtex/js-own,/home/schnack/tex/bibtex/js-mag,/home/schnack/tex/bibtex/js-mis,v20.bbl}



\end{document}